\documentclass[journal]{IEEEtran}
\usepackage[utf8]{inputenc}
\usepackage{authblk}
\usepackage{multirow}
\usepackage{graphicx}
\usepackage[outdir=./]{epstopdf}
\usepackage{epstopdf}
\usepackage{stmaryrd}
\usepackage{epsfig}
\usepackage{amsmath}
\usepackage{algorithmic}
\usepackage{mathtools}
\usepackage{array}
\usepackage{multirow}
\usepackage{stmaryrd}
\usepackage{multirow}
\usepackage{algorithmic}
\usepackage {algorithm}
\usepackage{amsthm,amssymb}
\usepackage{cite}
\usepackage{hyperref}

\usepackage[flushleft]{threeparttable}

\usepackage{xcolor}  % Blue font revision.
\usepackage{epstopdf}

\providecommand{\keywords}[1]
{
  \small	
  \textbf{\textit{Keywords---}} #1
}
\begin{document}

% \title{A Novel Wider Attention EfficientNet (WATT-EffNet) for Classifying Aerial Disaster Images}
\title{WATT-EffNet: A Lightweight and Accurate Model for Classifying Aerial Disaster Images}
% \title{A Novel Wider ATtention EfficientNET (WATENET) Network for Classifying Aerial Disaster Images}

\author{Gao Yu Lee, Tanmoy Dam, Md Meftahul Ferdaus, Daniel Puiu Poenar, and Vu N. Duong}

%\author{Gao Yu Lee\textsuperscript{\textsection},~\IEEEmembership{} Tanmoy Dam\textsuperscript{\textsection},~\IEEEmembership{} Md Meftahul Ferdaus, Daniel Puiu Poenar,~Vu N. Duong~\IEEEmembership{}$\;$\thanks{Gao Yu Lee, Tanmoy Dam, Md Meftahul~Ferdaus, Daniel Puiu Poenar, and Vu N. Duong
%are with~Air Traffic Management Research Institute (ATMRI), Nanyang Technological University,
%Singapore (e-mail: \protect\href{mailto:mdmeftahul.ferdaus@ntu.edu.sg}{mdmeftahul.ferdaus@ntu.edu.sg};
%\protect\href{mailto:tanmoy.dam@ntu.edu.sg}{tanmoy.dam@ntu.edu.sg};
%\protect\href{@e.ntu.edu.sg}{zhijun001@e.ntu.edu.sg};
%\protect\href{@ntu.edu.sg}}}

\author{Gao Yu Lee, Tanmoy Dam, Md Meftahul Ferdaus, Daniel Puiu Poenar and Vu N. Duong
            \thanks{Gao Yu Lee, Tanmoy Dam, Md Meftahul Ferdaus, and Vu N. Duong are with the Air Traffic Management Research Institute (ATMRI), Nanyang Technological University, Singapore. (email: GAOYU001@e.ntu.edu.sg, tanmoy.dam@ntu.edu.sg, mdmeftahul.ferdaus@ntu.edu.sg, vu.duong@ntu.edu.sg)}

            \thanks{Gao Yu Lee and Daniel Puiu Poenar are with the school of Electrical and Electronic Engineering (EEE), Nanyang Technological University, Singapore. (email: GAOYU001@e.ntu.edu.sg, EPDPuiu@ntu.edu.sg)
        }
        }

\maketitle

\begin{abstract}
Incorporating deep learning (DL) classification models into unmanned aerial vehicles (UAVs) can significantly augment search-and-rescue operations and disaster management efforts. In such critical situations, the UAV's ability to promptly comprehend the crisis and optimally utilize its limited power and processing resources to narrow down search areas is crucial. Therefore, developing an efficient and lightweight method for scene classification is of utmost importance. However, current approaches tend to prioritize accuracy on benchmark datasets at the expense of computational efficiency. To address this shortcoming, we introduce the Wider ATTENTION EfficientNet (WATT-EffNet), a novel method that achieves higher accuracy with a more lightweight architecture compared to the baseline EfficientNet. The WATT-EffNet leverages width-wise incremental feature modules and attention mechanisms over width-wise features to ensure the network structure remains lightweight. We evaluate our method on a UAV-based aerial disaster image classification dataset and demonstrate that it outperforms the baseline by up to 15 times in terms of classification accuracy and $38.3\%$ in terms of computing efficiency as measured by Floating Point Operations per second (FLOPs). Additionally, we conduct an ablation study to investigate the effect of varying the width of WATT-EffNet on accuracy and computational efficiency.  Our
code is available at \url{https://github.com/TanmDL/WATT-EffNet}.

% Unmanned aerial vehicles (UAVs) embedded with a deep learning (DL) classification model can be utilized for search-and-rescue operations, as well as monitoring and mitigating the effects of natural disasters. In such cases, the UAV must quickly comprehend the crisis scenario to limit search areas. This ensures that the UAV's limited power and processing resources are used properly and effectively. To achieve such goal, an effective and lightweight method for scene classification must be developed. Existing approaches, on the other hand, prioritize achieving the highest possible accuracy on classification-based benchmark datasets while ignoring computationally efficient observations. To mitigate such research gap, we have introduced a novel Wider ATTention EFFicientNet (WATT-EffNet) to achieve higher accuracy with a lighter architecture than the baseline EfficientNet. Implementation of width-wise incremental feature module and attention mechanism over width-wise features in WATT-EffNet contributes to the network's light structure. UAV-based aerial disaster image classification dataset is considered for analysing the effectiveness of the purposed WATT-EffNet. Our method surpassed baseline by up to a factor of 15 in terms of classification accuracy and by 38.3\% in terms of computing efficiency measured by the Floating Point Operations per second (FLOPs). An ablation study is also conducted by varying the width of WATT-EffNet to examine a higher accuracy at a lower FLOPs. 
%Our proposed method implementation is available on the \ref{}.
\end{abstract}

\keywords{ \textbf{Convolutional Neural Networks (CNN)}, \textbf{WATT-EffNet}, \textbf{Disaster Scene classification}, \textbf{Unmanned Aerial Vehicles (UAVs)}}

\section{Introduction}
\begin{figure}[hbt!]
    \centering
    \includegraphics[scale=0.42]{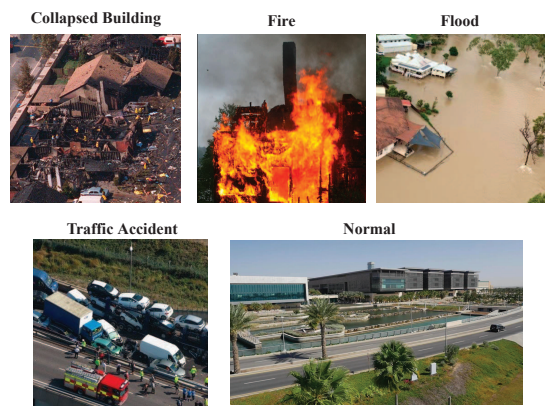}
    \caption{Example of an image from each respective disaster classes in the AIDER dataset.}
    \label{fig:image_sample}
\end{figure}

Recent technological advancements in unmanned aerial vehicles (UAVs) have significantly improved their capabilities for activities such as remote sensing and visual geological surveying, leading to greater efficiency and effectiveness \cite{valsan2020unmanned}. In particular, UAVs play a critical role in search and rescue operations, where they can monitor disaster areas for damage and locate survivors. By classifying disaster scenes, UAVs can quickly determine the type of disaster that has occurred and focus their search efforts. This is vital given the limited power and memory resources of UAVs. As noted by \cite{kyrkou2020emergencynet}, the four primary types of disasters that can be found in aerial image databases for emergency response applications are fires, floods, building collapses, and traffic collisions. Fig. \ref{fig:image_sample} displays representative samples of each image type. The dataset used in this study consists of aerial images of various disaster classes, including fires, floods, building collapses, and traffic collisions, collected from multiple sources such as the internet and UAV platforms. To simulate real-world scenarios as accurately as possible, the dataset also contains a significant number of non-disaster images labeled as normal, as highlighted in the same study.

%study by \cite{kyrkou2020emergencynet}. 

The objective of this study is to create an efficient and lightweight deep learning classifier that can swiftly detect and classify various events using effective sensors and microprocessors. To this end, we have analyzed various architectures, including MobileNet, SqueezeNet, ShuffleNet, and EfficientNet. These models were designed with the objective of being lightweight by incorporating a range of techniques, including depth-wise separable convolutions, pointwise filters, group convolution, channel shuffling, and strategic scaling of depth, width, and resolution. Specifically, MobileNet models were created for mobile platform applications and utilized depth-wise separable convolutions instead of traditional convolutions to reduce the number of training parameters relative to conventional convolutional networks with the same depth.  

%% Once highlighetd in blue
As an illustration, MobileNetV1 \cite{howard2017mobilenets} was the first MobileNet architecture proposed, requiring only 4.2 million training parameters. This is in contrast to the VGG16 and GoogleNet architectures, which require 138 million and 6.8 million parameters, respectively \cite{liu2015very}. MobileNetV2, a subsequent architecture to MobileNetV1, introduced further modifications that significantly reduced the number of training parameters required from 4.2 million to 3.4 million \cite{sandler2018mobilenetv2}. Other instances include the SqueezeNet \cite{iandola1602squeezenet}, which utilized 1$\times$1 pointwise filters rather than the 3$\times$3 filters in traditional convolutional networks to reduce computational costs. The ShuffleNet \cite{ma2018shufflenet} introduced group convolution and adds a channel shuffling operation in the depth-wise convolutional layer for efficient computation. Another model called the EfficientNet \cite{tan2019efficientnet} scales the depth, width and resolution of the network architecture strategically to achieve both computational efficiency and effectiveness. Despite the aforementioned modifications and reduced structural complexity, the number of training parameters utilized in these lightweight models still amounts to millions. Consequently, these models are not particularly suitable for prolonged UAV operations due to their high FLOPs demand on the on-board CPU \cite{yao2023lightweight}.

\begin{figure*}[hbt!]
    \centering
    \includegraphics[height = 10.5cm, width=16.0cm]{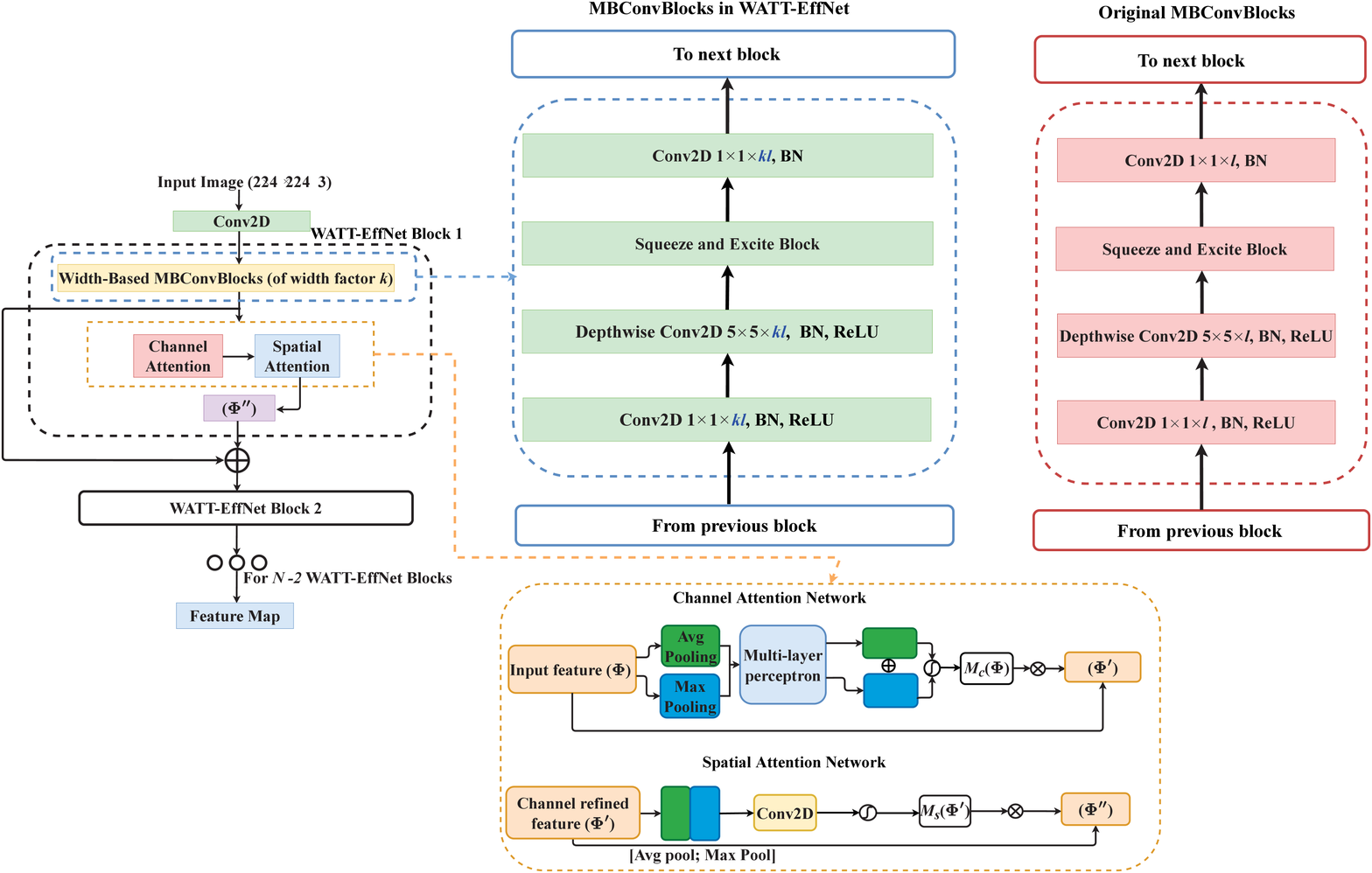}
    \caption{The algorithmic structure of our WATT-EffNet, as shown on the left of the figure. Our modification to the MBConv block layer using EfficientNet as the backbone is shown on the top right of the figure, as highlighted by the blue dotted box. We also illustrate the  original MBConv block layer for comparison (red dotted box). The attention mechanism architecture is illustrated in the dotted orange box on the bottom right of the figure.}
    \label{fig:overall_architecture}
\end{figure*}

Unmanned aerial vehicles (UAVs) are critical for finding survivors in disaster zones quickly and efficiently. However, the success of these missions heavily relies on the computational efficiency and effectiveness of the UAVs' on-board classification model. To address this challenge, we propose a novel Wider ATTENTION EfficientNet (WATT-EffNet) model that incorporates an attention module across a wider EfficientNet network. Unlike traditional models, this design emphasizes the width of the network rather than the depth. Additionally, by utilizing attention, it reduces the computational cost of the network by processing only key feature maps. The importance of increasing the width of the network while reducing its depth is supported by \cite{zagoruyko2016wide}, whose findings consistently show gains in classification accuracy for widened residual networks due to the increased representational power. Moreover, they have shown that this increase can be achieved with only a marginal increase in the number of training parameters, where a factor of 2 increase in the network's width with a depth reduction of 2.5 leads to a relatively smaller increase in the number of training parameters. As far as our knowledge extends, the amalgamation of a width-based architecture with an attention-based mechanism for UAV disaster scene classification in an existing EfficientNet model has not been attempted heretofore. In essence, the present letter outlines several contributions, which can be summarized as follows:

\begin{itemize}
    
    \item We present WATT-EffNet, an architectural innovation that leverages the principle of width as a foundation to augment the capabilities of the original EfficientNet model. Through the integration of attention modules and a reduction in overall complexity, this architecture aspires to surpass the performance of the original EfficientNet by promoting both efficiency and effectiveness. 
    \item WATT-EffNet architecture endeavors to strike a delicate balance between computational efficiency and effectiveness, a critical consideration in the context of classifying disaster scenes captured by UAVs, where mission success probability must be optimized amidst limited resource constraints. This approach deviates from the prevalent paradigm among state-of-the-art models, which do not necessarily take such operational constraints into account.
    \item The efficacy of our proposed WATT-EffNet architecture has been rigorously evaluated through the utilization of a subset of the AIDER dataset, where the results demonstrate that our model is capable of achieving superior $F_1$ scores, in contrast to the established baselines, while concurrently utilizing a substantially lower number of FLOPs, thereby exemplifying its computational efficiency.
\end{itemize}

\section{WATT-EffNet}

\subsection{Width-varying feature module}

%\textcolor{red}{(I'm not really sure if we have addressed reviewer II point 6 on giving a simple introduction to the whole and then give a detailed introduction to the different parts ?)}

The advent of deeper neural network architectures such as ResNet was a response to the problem of vanishing gradients, an occurrence that plagues the training of deep neural networks. The phenomenon of vanishing gradients is characterized by the diminution of the magnitudes of gradients computed during backpropagation, making it hard to make meaningful updates to the weights of the network. This, in turn, can impede the successful training of deep neural networks.  In order to overcome the problem of vanishing gradients, the concept of skip connections was introduced. These connections enable the gradients to circumvent one or more layers and be directly propagated to the shallower layers of the network, thus enabling the gradients to traverse the network with greater ease and fluidity. This mechanism promotes the training of deep neural networks by preserving the gradient magnitudes.
 For a given residual unit $i$, its input $Y_{i-1}$, the learnt feature mapping $\phi_{i}(\cdot)$, and the trainable parameters of the unit $\boldsymbol{\omega}_{i}$, the output $Y_{i}$ of the residual unit can be defined recursively as

\begin{equation}
Y_{i} = \phi_{i}(Y_{i}, \boldsymbol{\omega}_{i}) +Y_{i-1}.
\end{equation}

$\phi_{i}(\cdot)$ is often comprised of two to three stacked convolutional stages, which, apart from the convolution layers, also comprised of the batch normalization and a  Rectifier Unit (ReLU) as an activation function. In each of the convolutional layers, $\boldsymbol{\omega}_{i}$ comprised of the kernel size $n$ $\times$ $m$ and the filter size $l$, and hence $\boldsymbol{\omega}_{i} = \boldsymbol{\omega}_{i}(n,m,l)$, where the total parameter in the layer is $nml$.

To enable a wider residual unit, a widening factor $k$ is introduced in which any unit with $k > 1 $ are categorized as wide, with the original residual unit categorized as $k$ = 1. Since we are utilizing the EfficientNet as the backbone as mentioned in the introduction, the corresponding units modified is the MBConv Block ($d$). As illustrated in the top right of Fig. \ref{fig:overall_architecture}, the original MBConv block is composed of a 1$\times$1 convolutional layer followed by a 5$\times$5 depth-wise convolution. The Squeeze and Excite (SE) \cite{jie2018squeeze} block is then applied to improve feature representation by accounting for the inter-dependencies across features from different channels. Finally, another 1$\times$1 convolutional layer is applied. Except for the SE block, batch normalization and ReLU are incorporated in each layer of the MBConv block. In our WATT-EffNet, we expand upon the width $k$ in the MBConv block so that for each layer, $\boldsymbol{\omega}_{i} = \boldsymbol{\omega}_{i}(n,m,kl)$ (as shown in Fig. \ref{fig:overall_architecture}) and hence for block $i$

\begin{equation}
Y_{i} = \Phi_{i}(Y_{i}, \boldsymbol{\omega}_{i} (n,m,kl)) + Y_{i-1}
\label{eq:resnet_1}
\end{equation}

where $\Phi_{i}(\cdot)$ is now the feature mapping associated with the MBConv blocks. By applying (\ref{eq:resnet_1}) using one substitution steps, the forward pass for our WATT-EffNet can be expanded as

\begin{equation}
\begin{split}
Y_{i+2} = \Phi_{i+2}(Y_{i+1}, {\boldsymbol{\omega}}_{i+2}(n,m,kl)) +  Y_{i+1} \\
= \Phi_{i+1}(Y_{i} +{\boldsymbol{\omega}}_{i+1}(n,m,kl)) + \\ \Phi_{i+2}(Y_{i} + \Phi_{i+1}(Y_{i} + {\boldsymbol{\omega}}_{i+1}(n,m,kl)), {\boldsymbol{\omega}}_{i+2}(n,m,kl)).
\end{split}
\end{equation}

This forms the width-based MBConv block layer in each block of our WATT-EffNet as illustrated by the light-blue box in the Fig. \ref{fig:overall_architecture}. The depth of the baseline EfficientNet architecture has a total of 19 layers, including 17 MBConv blocks \cite{tan2019efficientnet}. We modified the number of MBConv blocks $d$ and $k$ in our approach such that the number of training parameters did not exceed 1M. Therefore, the possible combination of our WATT-EffNet architecture is represented as WATT-EffNet-$d$-$k$. More details about the variation in our architecture are given later in the experimental results section.

\subsection{Attention mechanism}

The attention module comprises the channel attention and the spatial attention modules, and their algorithmic structure is illustrated in the bottom right of Fig. \ref{fig:overall_architecture}. For the channel attention, the process involved extracting and squeezing the spatial dimension of the input feature map $\boldsymbol{\Phi}$, which is extracted from the wider structure network in the previous subsection, followed by parallel average and max pooling. The output feature from each pooling are then fed into a shared Multi-Layer Perceptron (MLP) network, with the resultant features finally merged using element-wise summation. The mathematical representation can be described in the following equation.

\begin{equation} \label{eq1}
\begin{split}
\boldsymbol{M_{c}(\boldsymbol{\Phi})} &= \sigma(MLP(AvgPool(\boldsymbol{\Phi}) + MLP(MaxPool(\boldsymbol{\Phi}))) \\
& = \sigma(\boldsymbol{\Omega_{1}}(\boldsymbol{\Omega_{0}}(\boldsymbol{{\Phi^{c}}_{avg}})) +\boldsymbol{W_{1}}(\boldsymbol{\Omega_{0}}(\boldsymbol{{\Phi^{c}}_{max}}))),
\end{split}
\end{equation}

where $\sigma$ denotes the sigmoid activation function, $\boldsymbol{\Omega_{0}}$, $\boldsymbol{\Omega_{1}}$ are the weights associated with the MLPs, $\boldsymbol{M_{c}(\boldsymbol{\Phi})}$ is the attention map associated with the channel attention, and $\boldsymbol{{\Phi^{c}}_{avg}}$ and $\boldsymbol{{\Phi^{c}}_{max}}$ denotes the feature maps obtained from the average and max pooling, respectively. For the spatial attention, both average pooling and max pooling are applied once again and concatenated along the channel axis. A convolutional layer of filter size 7$\times$7 ($F^{7\times7}$) is then applied to generate the spatial attention map. Therefore, the spatial feature-map can be described in the following form,

\begin{equation} \label{eq2}
\begin{split}
\boldsymbol{M_{s}(\Phi)} &= \sigma(F^{7\times7}([AvgPool(\boldsymbol{\Phi}); MaxPool(\boldsymbol{\Phi})])) \\
&= \sigma(F^{7\times7}([\boldsymbol{{\Phi^{s}}_{avg}}; \boldsymbol{{\Phi^{s}}_{max}}])),
\end{split}
\end{equation}

where $\boldsymbol{M_{s}(\Phi)}$ is the attention map associated with the spatial attention and $\boldsymbol{{\Phi^{s}}_{avg}}$ and $\boldsymbol{{\Phi^{s}}_{max}}$ denotes the feature map obtained from the average and max pooling respectively in this part of the attention module. Finally, the overall attention mechanism is described as follows, 

\begin{equation} \label{eq3}
\begin{split}
\boldsymbol{\Phi'} &= \boldsymbol{M_{c}(\Phi)} \otimes \boldsymbol{\Phi}, \\
\boldsymbol{\Phi''} &= \boldsymbol{M_{s}(\Phi')} \otimes \boldsymbol{\Phi'}, \\
\end{split}
\end{equation}

where $\otimes$ denotes the outer product between the relevant feature and attention maps. This forms the attention block component of a WATT-EffNet block as illustrated by the orange box in Fig. \ref{fig:overall_architecture}. A skip connection is also performed before the next WATT-EffNet block. Therefore, the proposed architecture can be repeated and used as a single modular block. This means that the proposed design can be repeated multiple times to create a deeper network, while still maintaining the benefits of the skip connections and other features that have been discussed. This modular approach allows for flexibility and scalability in the design of the network.

\section{Experimental Details}

%\textcolor{red}{(In subsequent sections, the table are placed a little bit behind, to address point number 4 raised by reviewer II.)}

WATT-EffNet is evaluated against SOTA methods using a subset of the AIDER dataset, which serves as a benchmark\cite{kyrkou2020emergencynet}. We compare methods that use minimal training parameters for algorithmic efficiency. Therefore, we did not include models like ResNet50 \cite{he2016deep}, VGG16, and Xception \cite{chollet2017xception} which require more than 5 million training parameters. Instead, we include MobileNetV1 \cite{howard2017mobilenets}, MobileNetV2 \cite{sandler2018mobilenetv2}, SqueezeNet \cite{iandola1602squeezenet}, ShuffleNet \cite{ma2018shufflenet}, EfficientNet \cite{tan2019efficientnet} and EmergencyNet \cite{kyrkou2020emergencynet}. The subset of the AIDER dataset used preserves class imbalance distributions. The train-valid-test-split ratio is 4:1:2 as same in the AIDER experiment. The original dataset has 700 images per disaster class and 5700 normal images, while the subset has 6433 images, as illustrated in Table \ref{table:data_class_distribution}.

\begin{table}[hbt!]
    \centering
    \caption{List of training, validation and test image sets for each class in the subset of the AIDER dataset. }
    \vspace{0.3mm} % Adjust the height of the space between caption and tabular
\begin{tabular}{c|c|c|c|c}
 \hline
 \textbf{Class} & \textbf{Train} & \textbf{Valid} & \textbf{Test} & \textbf{Total per Class}\tabularnewline
 \hline
 Collapsed Building & 367 & 41 & 103 & 511\tabularnewline
 \hline
 Fire & 249 & 63 & 209 & 521\tabularnewline
  \hline
 Flood & 252 & 63 & 211 & 526\tabularnewline
  \hline
 Traffic & 232 & 59 & 194 & 485\tabularnewline
 \hline
Normal & 2107 & 527 & 1756 & 4390\tabularnewline
\hline
\textbf{Total Per Set} & \textbf{3207} & \textbf{753} & \textbf{2473} & \textbf{6433}\tabularnewline
 \hline
\end{tabular}
\label{table:data_class_distribution}
\end{table}

All images were resized to 224$\times$224$\times$3. To address class imbalance, we applied under-sampling to the training and validation sets using the RandomUnderSampler module from the imblearn library, as in the AIDER experiment. All simulations were done using the Tensorflow Keras library in Python on Google's Colab Pro+ Tesla T4 GPUs and TPUs. All the experimental simulations are evaluated (including our approach) with an epoch set to 300. In the preprocessing stage, the intensity values of each pixel in the images were divided by 255 to normalize the data. Additionally, a kernel regularizer with a coefficient of 1e-4 was incorporated into every convolutional block layer. Our approach employed both categorical cross-entropy and cosine loss as the training loss, as suggested by \cite{barz2020deep}, which argues that this combination is effective for models that are trained from scratch and without any pre-training on small data, which is the case for our dataset. Both losses were given equal importance in the combined loss function. The optimization algorithm used was Root-Mean-Square propagation (RMSprop) and was set to a decay rate of 1e-6.

The development of our proposed WATT-EffNet model is predicated on the utilization of the wider structure and attention mechanism principle, thus the identification of the optimal width to depth ratio ($k$ and $d$ respectively) is of paramount importance. This necessitates the examination of a plethora of possible permutations and combinations of the width ($k$) and depth ($d$) utilizing a range of values for $k = \{2, 3, 4, 5, 6, 7 \}$ and $d = \{1, 3, 5\}$. The next section of this research will be dedicated to the exploration of the aforementioned permutations and combinations through the utilization of the AIDER dataset, in order to arrive at the optimal values for $k$ and $d$.

\section{Results and Discussions}

% The performance of the proposed WATT-EffNet ($d$=3, $k$=6) is compared with several state-of-the-art (SOTA) models in Table \ref{tab:comapred_baseline}. The considered SOTA models are MobileNetV1, MobileNetV2, SqueezeNet, ShuffleNet, EfficientNet, and EmergencyNet. The comparison is based on three widely used performance metrics, namely $F_1$ score, number of floating-point operations (FLOPs), and number of parameters.

% Table \ref{tab:comapred_baseline} compares the performance of several SOTA models along with the proposed WATT-EffNet($d$=3, $k$=6). The SOTA models include MobileNetV1, MobileNetV2, SqueezeNet, ShuffleNet, EfficientNet and EmergencyNet. The three commonly used performance metrics are $F_1$ score, FLOPs, and number of parameters. 

Table \ref{tab:comapred_baseline} illustrates that the proposed model has demonstrated a significant improvement in performance in comparison to the baseline EfficientNet model as well as the various benchmark methods. Specifically, the performance improvement has been quantified as a $10.6\%$ increase in terms of $F_1$ score. Additionally, the proposed model has also shown a drastic reduction in computational complexity as it requires 35 times fewer FLOPs when compared to the experimental conditions. Furthermore, it is worth noting that although the EmergencyNet model utilizes fewer parameters, the WATT-EffNet model's wider structure design allows for more efficient utilization of parameters, resulting in lower FLOPs and overall computational complexity. The results of these evaluations have revealed a substantial enhancement in performance as compared to the EmergencyNet model with an improvement of more than $6.5\%$. Furthermore, the proposed model also demonstrates a remarkable reduction in computational complexity, as quantified by FLOPs, with a 8-fold decrease in comparison to the EmergencyNet model.

\begin{table}[htp]
   \centering
 \caption{$F_1$ scores, FLOPs, and training parameters for the existing SOTA CNN-based approaches on the AIDER  dataset$^*$. Bolded values denote the highest $F_1$ score, lowest FLOPs and lowest training parameters.}
    \vspace{0.3mm} % Adjust the height of the space between caption and tabular
\begin{tabular}{c|c|c|c}
 \textbf{SOTA Model} & \textbf{$F_{1} (\%)$}($\shortuparrow$) & \textbf{FLOPs} ($\shortdownarrow$) & \textbf{Parameters} ($\shortdownarrow$) \tabularnewline
\hline 
MobileNetV1 \cite{howard2017mobilenets}  & 80.2 & 972 & 3,233,861 \tabularnewline
\hline 
MobileNetV2 \cite{sandler2018mobilenetv2} & 81.0 & 625 & 2,282,629 \tabularnewline
\hline 
SqueezeNet \cite{iandola1602squeezenet} & 82.3 & 531 & 725,073  \tabularnewline
\hline 
ShuffleNet \cite{ma2018shufflenet} & 81.7 & 972 & 4,023,865  \tabularnewline
\hline 
EfficientNet  \cite{tan2019efficientnet} & 80.0 & 774 & 3,499,453 \tabularnewline
\hline 
EmergencyNet \cite{kyrkou2020emergencynet}  & 83.1 & 185 & \textbf{94,420}  \tabularnewline
\hline 
\textbf{WATT-EffNet-3-6} & \textbf{88.5} & \textbf{22} & 688,661  \tabularnewline
\hline 
\end{tabular}
\label{tab:comapred_baseline}
 \begin{tablenotes}
    \item *All models are trained on the same environment.
    \end{tablenotes}
\end{table}

%%%%%%%%%

\begin{figure}[hbt!]   %confusionmatrix1
    \includegraphics[height = 6.5cm, width=8.0cm]{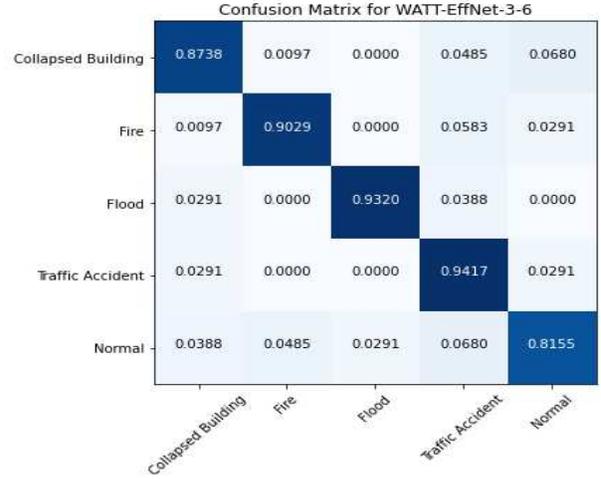}
    \caption{The confusion matrix for our WATT-EffNet-3-6.}
    \label{fig:confusion_mat}
\end{figure}

\begin{figure}[hbt!]
    \centering
    \includegraphics[height = 5.2cm, width=6.6cm]{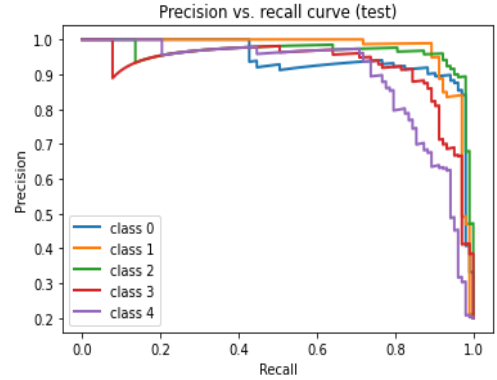}
    \caption{The precision vs recall curves for the predicted classes using our aforementioned model against the original classes. Here, class 0-5 represents collapsed building, fire, flood, traffic incident and the normal class respectively.}
    \label{fig:prec_rec}
\end{figure}

We have examined the performance of each class through a confusion matrix and a precision vs recall curve, which is illustrated in Fig. \ref{fig:confusion_mat} and \ref{fig:prec_rec} respectively. Our analysis revealed that the normal class had the lowest prediction percentage (81.6\%) among all the classes due to the presence of images that did not have an ideal view perspective and the key features only occupied a small area of the image. On the other hand, the traffic incident class had the highest prediction percentage (94.2\%) among all the classes. However, the $F_1$ values obtained in our work were lower than those obtained in the EmergencyNet. This is because the methods utilized in our context are mostly, if not all, CNN-based network, and such network usually perform better with increasing amount of training data. Nevertheless, our optimal lightweight algorithmic design is still capable of maximizing learning even when using a limited dataset for training. Lastly, it is worth mentioning that some works such as \cite{he2022generating} and \cite{mirik2012utility} have highlighted the significance of incorporating a finer spatial resolution during classification and have assessed its impact on the classification performance. This is crucial in disaster scenarios, as they typically occur within a limited region in a larger captured field of view. While this aspect was not the primary focus of our research, it is an intriguing and critical direction that we plan to explore in our future works.

% {\textcolor{red}{(Addressing reviewer I point 3 and reviewer II point 7, but one reference from each of what they give.)}

%We have examined of the each class performance through confusion matrix and also illustrated in Fig. \ref{fig:confusion_mat} and \ref{fig:prec_rec}. Our analysis of the confusion matrix revealed that the normal class had the lowest prediction percentage (81.6\%) among all the classes due to the presence of images that did not have an ideal view perspective and that the key features only occupied a small area of the image. On the other hand, the traffic incident class had the highest prediction percentage (94.2\%) among all the classes. However, the $F_1$ values obtained in our work were lower than those obtained in \cite{kyrkou2020emergencynet}. 

%Hence, we expect that if the full AIDER dataset were available and used instead for training our network, it would yield even higher performance metrics value. Since our method have yielded higher F1 scores than the EmergencyNet in the AIDER subset, such a better performance trend is then to be expected for the full dataset.

%%%%%%%%%%%%

\begin{table}[htp]
   \centering
 \caption{$F_1$ scores, FLOPs, training parameters for our WATT-EffNet framework as a function of number of MBConv Blocks \textbf{(first value in model variant name)} and widths \textbf{(second value in model variant name)} with respect to the original EfficientNet framework$^*$. }
    \vspace{0.3mm} % Adjust the height of the space between caption and tabular
\begin{tabular}{c|c|c|c}
 \textbf{Model Variant} & \textbf{$F_{1} (\%)$}($\shortuparrow$) & \textbf{FLOPs} ($\shortdownarrow$) & \textbf{Parameters} ($\shortdownarrow$) \tabularnewline
\hline 
WATT-EffNet-1-2 & 63.1$\pm$0.02 & 22 & 8,501 \tabularnewline
\hline 
WATT-EffNet-1-3 & 64.2$\pm$0.07 & 22 & 13,693 \tabularnewline
\hline 
WATT-EffNet-1-4 & 66.5$\pm$0.38 & 22 & 19,909 \tabularnewline
\hline 
WATT-EffNet-1-5 & 68.2 $\pm$0.03 & 22 & 27,149 \tabularnewline
\hline 
WATT-EffNet-1-6 & 70.4 $\pm$0.87 & 22 & 35,413 \tabularnewline
\hline 
WATT-EffNet-1-7 & 72.7$\pm$0.05 & 22 & 44,701 \tabularnewline
\hline 
WATT-EffNet-3-2 & 84.2$\pm$1.03 & 22 & 106,629  \tabularnewline
\hline 
WATT-EffNet-3-3 & 83.3$\pm$1.08 & 22 & 205,673  \tabularnewline
\hline 
WATT-EffNet-3-4 & 86.0$\pm$1.05 & 22 & 335,693   \tabularnewline
\hline 
WATT-EffNet-3-5 & 85.2$\pm$0.98  & 22 & 496,689   \tabularnewline
\hline 
\textbf{WATT-EffNet-3-6} & \textbf{88.5$\pm$0.76} & \textbf{22} & \textbf{688,661}   \tabularnewline
\hline 
WATT-EffNet-3-7 & 87.3$\pm$0.72 & 22 & 911,609  \tabularnewline
\hline 
WATT-EffNet-5-2 & 86.8$\pm$1.05 & 22 & 371,493   \tabularnewline
\hline 
WATT-EffNet-5-3 & 86.4$\pm$0.93 & 22 & 720,233   \tabularnewline
\hline 
\end{tabular}
\label{tab:final_results}
 \begin{tablenotes}
    \item *All models are trained on the same environment.
    \end{tablenotes}
\end{table}

%Table III depicts the F1 scores, FLOPs and parameters utilized for the SOTA approaches utilized in \cite{kyrkou2020emergencynet}. All the approaches are trained from scratch on the AIDER subset. Indeed, our WATT-EffNet-3-6 approach in this subset exceeded that of all the SOTA in the F1 scores, especially as relative to EmergencyNet using lower FLOPs, in spite of utilizing a relatively larger number of training parameters (688,661).

%\textcolor{red}{(Addressing reviewer II point 5)}

\subsubsection*{Ablation Studies}

The WATT-EffNet model was developed with the goal of achieving high accuracy while maintaining a lightweight architecture. To evaluate its performance, we conducted an ablation study by varying the number of MBConv Blocks ($d$) and width ($k$) of the model. Additionally, we extended the study to include variations with and without the attention mechanism, as recorded in Table \ref{tab:final_results} and \ref{tab:final_results_noat} respectively. 

%%%%chatgpt generated 
The results of various WATT-EffNet model variants considering attention mechanism are presented in Table \ref{tab:final_results} and are quantified by the standard deviation and mean $F_1$ scores. The first number in the variant name indicates the number of MBConv blocks ($d$) used in the network, while the second number represents the width multiplier ($k$). The $F_1$ score, FLOPs, and number of parameters are reported for each variant. The best results were obtained when $d$ was equal to 3 and $k$ was equal to 6, however it is noteworthy that the results were similar or even better when using a smaller value of $k=2$ with a smaller number of parameters than the SOTA models. It is worth mentioning that we examined all the possible conditions until the parameters did not exceed 1M. 

Similarly, Table \ref{tab:final_results_noat} presents the results of various WATT-EffNet model variants where the attention mechanism was discarded. The $F_1$ score, FLOPs, and number of parameters are reported for each variant. The best results were obtained when $d$ was equal to 3 and $k$ was equal to 6. The $F_1$ scores obtained in Table \ref{tab:final_results_noat} are lower than those given in Table \ref{tab:final_results}. This indicates that incorporating attention improves the classification performance by a significant margin in some cases and by a small margin in others (e.g., WATT-EffNet-3-3 displayed an improvement in $F_1$ scores by only around 0.2\% while WATT-EffNet-1-6 demonstrated an improvement by around 6\%). It is worth mentioning that the FLOPs used remains the same regardless of whether attention is incorporated or not, which implies that attention modules can be incorporated even for a given limited computational resource since no additional costs are incurred to enhance the classification performance.

\begin{table}[htp]
   \centering
 \caption{$WATT-EffNet$ metrics without the attention mechanism$^*$.}
    \vspace{0.3mm} % Adjust the height of the space between caption and tabular
\begin{tabular}{c|c|c|c}
 \textbf{Model Variant} & \textbf{$F_{1} (\%)$}($\shortuparrow$) & \textbf{FLOPs} ($\shortdownarrow$) & \textbf{Parameters} ($\shortdownarrow$) \tabularnewline
\hline
WATT-EffNet-1-2 & 62.1$\pm$0.70 & 22 & 8,501 \tabularnewline
\hline 
WATT-EffNet-1-3 & 63.2$\pm$0.25 & 22 & 13,693 \tabularnewline
\hline 
WATT-EffNet-1-4 & 63.9$\pm$0.28 & 22 & 19,909 \tabularnewline
\hline 
WATT-EffNet-1-5 & 65.9 $\pm$1.39 & 22 & 27,149 \tabularnewline
\hline 
WATT-EffNet-1-6 & 66.4 $\pm$2.44 & 22 & 35,413 \tabularnewline
\hline 
WATT-EffNet-1-7 & 69.1$\pm$0.91 & 22 & 44,701 \tabularnewline
\hline 
 WATT-EffNet-3-2 & 81.5$\pm$1.04 & 22 & 106,629\tabularnewline
  \hline
 WATT-EffNet-3-3 & 83.1$\pm$1.22 & 22 & 205,673\tabularnewline
 \hline
 WATT-EffNet-3-4 & 85.2$\pm$0.98 & 22 & 335,693 \tabularnewline
   \hline
 WATT-EffNet-3-5 & 85.0$\pm$0.92 & 22  & 496,689 \tabularnewline
 \hline
 \textbf{WATT-EffNet-3-6} &  \textbf{85.4$\pm$1.29} & \textbf{22}  & \textbf{688,661} \tabularnewline
  \hline
 WATT-EffNet-3-7 & 85.2$\pm$0.46 & 22  & 911,609\tabularnewline
 \hline
WATT-EffNet-5-2 & 83.5$\pm$0.88 & 22 & 371,493 \tabularnewline
\hline
WATT-EffNet-5-3 & 84.3$\pm$0.89 & 22 & 720,233 \tabularnewline
\hline
\end{tabular}
\label{tab:final_results_noat}
 \begin{tablenotes}
    \item *All models are trained on the same environment.
    \end{tablenotes}
\end{table}

% Table \ref{tab:final_results_noat} lists the corresponding F1 scores and FLOPs value variation when we did not incorporate the attention modules. The value underlined denote the 

\section{Conclusions}

The present study introduces a novel width-based EfficientNet architecture, named Wider ATTENTION EfficientNet (WATT-EffNet), that incorporates the convolutional block attention mechanism. This new network architecture enables the creation of a shallower EfficientNet-based classifier, resulting in reduced computational processing requirements, by selectively attending to only the most informative features present on the feature map. The proposed WATT-EffNet variants were trained and evaluated on a challenging subset of the AIDER dataset, featuring a highly imbalanced distribution of non-disaster and disaster images of different classes. The obtained results consistently demonstrate the effectiveness and efficiency of the proposed architecture in comparison to state-of-the-art methods. The significance of this work lies in highlighting the role of width and attention in a CNN-based network in enhancing both efficiency and classification performance, making it a promising solution for visual-based UAV search-and-rescue operations. In future works, the proposed WATT-EffNet architecture will be tested in real-time UAV operations for validation, and alternative methods such as GAN-based augmentation \cite{dam2020mixture, dam2022latent} will be explored to further improve the model's performance.

\section*{Acknowledgement}

This research is supported by the Civil Aviation Authority of Singapore and NTU under their collaboration in the Air Traffic Management Research Institute. Any opinions, findings and conclusions or recommendations expressed in this material are those of the author(s) and do not necessarily reflect the views of the Civil Aviation Authority of Singapore.

\bibliographystyle{IEEEtran}
\bibliography{ref}

\end{document}